\documentclass[amsthm]{elsart}

\usepackage{yjsco}
\usepackage{natbib}

\usepackage{amssymb}
  \usepackage{amsthm}

\begin{document}

\begin{frontmatter}

\title{A note on the five valued conjectures of Johansen and Helleseth and zeta functions}

\thanks{ 
This work is supported in part by 
the National Natural Science Foundation of China under Grants 61271222 and 60972033.}

 \author{Xiaogang Liu }\footnote{corresponding author}
\address{ Department of
Computer Science and Engineering, Shanghai Jiao Tong University, Shanghai 200240, P.R. China}
\ead{liuxg0201@163.com}

\author{Michael Harrison }
\address{ Computational Algebra Group, School of Mathematics and Statistics, The University of Sydney, Australia } 
\ead{michael.harrison@sydney.edu.au}

\author{Yuan Luo   }
\address{ Department of
Computer Science and Engineering, Shanghai Jiao Tong University, Shanghai 200240, P.R. China}
\ead{yuanluo@sjtu.edu.cn}

\begin{abstract}
For the complete five-valued cross-correlation distribution  between two $m$-sequences ${s_t}$ and ${s_{dt}}$ of period $2^m-1$ that differ by the decimation $d={{2^{2k}+1}\over {2^k+1}}$ where $m$ is odd and $\mbox{gcd}(k,m)=1$, Johansen and Hellseth expressed it in terms of some exponential sums. And two conjectures are presented that are of interest in their own right.
In this correspondence we study these conjectures for the particular case where $k=3$, and the cases $k=1,2$ can also be analyzed in a similar process. When $k>3$, the degrees of the relevant polynomials will become higher. 
 Here the multiplicity of the biggest absolute value of the cross-correlation is  no more than one-sixth of the multiplicity corresponding the smallest absolute value. 
\end{abstract}


\end{frontmatter}

\section{Introduction} \label{sec1}

Let ${s_t}$ be an $m$-sequence of period $2^m-1$ and ${s_{dt}}$ be the decimated sequence of the same period with  $\mbox{gcd}(d,2^m-1)=1$.
The cross-correlation function between these two sequences at shift $\tau$ is defined by
\[
C_d(\tau)=\sum\limits_{t=0}^{2^m-2}(-1)^{s_{t+\tau}+s_{dt}}
\]
where $0\leq \tau < 2^m-1$.

Let $\mbox{Tr}_m(x)$ be the trace mapping from $\mbox{GF}(2^m)$ to $\mbox{GF}(2)$.
Then the $m$-sequence can be represented as $s_t=\mbox{Tr}_m(\alpha^t)$, and we can write   the cross correlation as
\[
C_d(\tau)=\sum\limits_{x\in \mbox{GF}(2^m)^*}(-1)^{\mbox{Tr}_m(ax+x^d)}
\]
where $a=\alpha^{\tau}$, $\alpha$ is a primitive element of the finite field $\mbox{GF}(2^m)$, and $\mbox{GF}(2^m)^*=\mbox{GF}(2^m)\backslash \{0\}$.

It has been important to determine the value distribution of the cross-correlation between two $m$-sequences \citet{H001,HK001,N001}. 
Johansen and Hellseth achieved  the five-valued cross-correlation distribution for a decimation $d={{2^l+1}\over {2^k+1}}$ where $k=1$ and $l=2$, i.e., $d=5/3$ \citet{JH001} . It seems that the correlation distribution is the same for  any $k$ and $l=2k$ when $\mbox{gcd}(k,m)=1$l. In \citet{JHK001}, the correlation distribution is given in terms of some exponential sums, and two conjectures are presented.  
 Proving these conjectures would imply that the correlation distribution is the same as for the case $k=1$, and they proved the case $k=2$. In this note we study the case $k=3$. Boston and McGuire considered the weight distributions of binary cyclic codes with at most five nonzero weights \citet{BM001}.
For relevant studies of cyclic codes, please refer to \citet{DL001,DY001,FL001,LL001,LTW001,X001,YC001,ZDLZ001}.

In the following, Section \ref{secII} recalls that the complete five-valued distribution can be reduced to computing the number $A_1$ of solutions of a special equation system. Furthermore, some exponential sums and two conjectures are presented. Section \ref{secIII} studies those two conjectures for $k=3$.

\section{Preliminaries} \label{secII}

In this section, relevant knowledge about the five-valued cross-correlation distribution is presented. Several kind of exponential sums and two conjectures about them are recalled. Theorems \ref{thm001} and \ref{thm002} are for the distribution in the general case, which are expressed in terms of the exponential sums. They originated when determining the number of solutions of the following system of euqations.
\begin{defn}\label{def02}
Let $A_a$, where $a\in \mbox{GF}(2^m)$, denote the number of solutions of the following equation system:
\[
\begin{array}{ll}
&x+y+z+u=a\\
&x^{2^k+1}+y^{2^k+1}+z^{2^k+1}+u^{2^k+1}=0\\
&x^{2^l+1}+y^{2^l+1}+z^{2^l+1}+u^{2^l+1}=0.\\
\end{array}
\]
\end{defn}
The notation $A_a=|E_{k,l}(a,0,0)|$ depends on $k$ and $l$, but these values will be clear from the context so we will usually omit $k$ and $l$ as indices.

There are the following important exponential sums \citet{JHK001}:
\[
\begin{array}{ll}
K_m &=\sum\limits_{x\in \mbox{GF}(2^m)^*}(-1)^{\mbox{Tr}_m(x+x^{-1})} \\
C_m &=\sum\limits_{x\in \mbox{GF}(2^m)}(-1)^{\mbox{Tr}_m(x^{2^k+1}+x)} \\
G_m^{(k)} &=\sum\limits_{x\in \mbox{GF}(2^m)^*}(-1)^{\mbox{Tr}_m(x^{2^k+1}+x^{-1})}.
\end{array}
\]
The first exponential sum $K_m$ is the celebrated Kloosterman sum. The second sum $C_m$ apparently depends on $k$, but it follows from Lahtonen, McGuire, and Ward \citet{LMW001} that
\[
\begin{array}{ll}
C_m&=+2^{(m+1)/2}, \mbox{if} \ m=\pm 1 \ \ (\mbox{mod} \ 8) \\
   &=-2^{(m+1)/2}, \mbox{if} \ m=\pm 3 \ \ (\mbox{mod} \ 8) \\
\end{array}
\]
when $m$ is odd and $\mbox{gcd}(k,m)=1$. 

The complete settlement of the correlation distribution for the case $d={{2^{2k}+1}\over {2^k+1}}$ depends on the following two conjectures. Together with the results in  \citet{JHK001} will show that the solution is the same as for $d=5/3$ which is determined by \citet{JH001}.

\begin{conj}\label{conj001}
For any positive integer $k$
\[
G_m^{(k)}=G_m^{(\mbox{gcd}(k,m))}
\]
i.e., $G_m^{(k)}$ depends only on $\mbox{gcd}(k,m)$. In particular, when $\mbox{gcd}(k,m)=1$ then
\[
G_m^{(k)} =\sum\limits_{x\in \mbox{GF}(2^m)^*}(-1)^{\mbox{Tr}_m(x^{3}+x^{-1})}.
\]
\end{conj}

Another important exponential sum presented in \citet{JHK001} is given in the following.

\begin{defn}\label{def01}
Let $m$ be odd, $\mbox{gcd}(k,m)=1$, and
\[
f(v)={{(v^{2^k}+1)v^{2^k}}\over {(v^{2^k}+v)^{2^k+1}}}
\]
where we define $f(0)=f(1)=0$ and
\[
K_m' =\sum\limits_{x\in \mbox{GF}(2^m)^*}(-1)^{\mbox{Tr}_m(f(v))}.
\]
\end{defn}

\begin{conj}
\[
K_m'=K_m.
\]
\end{conj}

The exponential sums $K_m, C_m$ and $G_m^{(1)}=G_m$ are needed because of the following result, the more general cases of which can be found in \citet{ZHC001}.

\begin{lem}
Let $m\geq 5$ be an odd integer. The number of solutions $A_1=|E_{1,2}(1,0,0)|$ of the system of equations
\[
\begin{array}{ll}
&x+y+z+u=1\\
&x^3+y^3+z^3+u^3=0\\
&x^5+y^5+z^5+u^5=0\\
\end{array}
\]
where $x,y,z,u\in \mbox{GF}(2^m)$, is given by
\[
A_1=2^m+1+3G_m-2K_m-2C_m
\]
where $G_m=G_m^{(1)}$.
\end{lem}

Johansen and Helleseth gave the following result about the cross-correlation distribution between two $m$-sequences \citet{JH001}.

\begin{thm} [Johansen et al.(2009)] \label{thm001}
Let $d={{2^{2k}+1}\over {2^k+1}}$ where $m$ is odd and let $\mbox{gcd}(k,m)=1$. Then $C_d(\tau)$ takes on the following five values $-1,-1-2^{(m+1)/2},-1+2^{(m+1)/2},-1-2^{(3m+1)/2},-1+2^{(3m+1)/2}$ with multiplicities $N_0,N_{-1},N_1,N_{-2},N_2$, respectively. Let $A_1=|E_{k,2k}(1,0,0)|$. Then the cross correlation distribution is as follows.
\begin{enumerate}
\renewcommand{\labelenumi}{$($\mbox{\roman{enumi}}$)$}
\item
In the case $\mbox{gcd}(3,m)=1$, we have the distribution
\[
\begin{array}{ll}
N_{-2}=N_2={A_1\over {96}}\\
N_{-1}={{3\cdot 2^{m+1}-3\cdot 2^{(m+3)/2}-A_1}\over {24}}\\
N_{1}={{3\cdot 2^{m+1}+3\cdot 2^{(m+3)/2}-A_1}\over {24}}\\
N_0=2^{m-1}-1+{A_1\over {16}}.
\end{array}
\]
\item
In the case $\mbox{gcd}(3,m)=3$, we have the distribution
\[
\begin{array}{ll}
N_{-2}={{-3\cdot 2^{(m+5)/2}+A_1}\over {96}}\\
N_2={{3\cdot 2^{(m+5)/2}+A_1}\over {96}}\\
N_{-1}=N_1={{3\cdot 2^{m+1}-A_1}\over {24}}\\
N_0=2^{m-1}-1+{A_1\over {16}}.
\end{array}
\]
\end{enumerate}
\end{thm}

For the case $k=1$, Johansen and Helleseth calculated the exponential sums $K_m,C_m$ and $G_m^{(1)}$ explicitly  \citet{JH001} . Hence, they provided the complete cross-correlation distribution for the decimation $d=5/3$ by determing $A_1=|E_{1,2}(1,0,0)|$.

Under the condition that $m$ is odd, later they found a similar expression for $A_1=|E_{k,2k}(1,0,0)|$ for any $k$ when $\mbox{gcd}(k,m)=1$ \citet{JHK001}.
 They showed that $A_1=|E_{k,2k}(1,0,0)|$ can be expressed in terms of $K_m',C_m$ and $G_m^{(k)}$.

\begin{thm}[Johansen et al.(2009)] \label{thm002}
Let $A_1=|E_{k,2k}(1,0,0)|$ where $m$ is odd and $\mbox{gcd}(k,m)=1$. Then
\[
A_1=2^m+1+3G_m^{(k)}-2K_m'-2C_m.
\]
\end{thm}

Under the assumption of the two conjectures, the following corollary holds, and they proved the conjecture for the case  $k=2$ \citet{JHK001}.

\begin{cor}\label{cor001}
If Conjectures 1 and 2 hold, then the correlation distribution for $d={{2^{2k}+1}\over {2^k+1}}$ is independent of $k$ when $m$ is odd and $\mbox{gcd}(k,m)=1$.

\end{cor}

\section{Main results}\label{secIII}

We investigate the two conjectures separately in the following two subsections. Before this let's recall some facts about zeta functions and Newton's identities.

For a positive integer $s$ , let $N_s$ denote the number of $\mathbb{F}_s$-rational points on the projective curve corresponding to homogeneous polynomial $f(x,y,z)$ and define the zeta function of the curve by
\[
\mathcal{Z}(t)=\mbox{exp}\{\sum\limits_{s=1}^{\infty}(N_s/s)t^s\},
\]
the series being convergent for $|t|\leq q^{-t}$ by trivial estimate of $N_s$. Here $\mathbb{F}$ is a finite field with $q$ element, and $\mathbb{F}_s$ is an extention of $\mathbb{F}$ with degree $s$. Then  Weil has shown that zeta function $\mathcal{Z}(t)$ is in fact a rational function of the form \citet{LH001}

\[
\mathcal{Z}(t)={{L(t)}\over{(1-t)(1-qt)}}
\]
where $L(t)$ is a polynomial of degree $2g$ with integer coefficients and constant term $1$, and $g$ is the genus. If one writes
\[
L(t)=\prod\limits_{j=1}^{2g}(1-\omega_jt),
\]
then
\[
N_s=q^s+1-\sum\limits_{j=1}^{2g}\omega_j^{s} \ \ \mbox{for all} \ \ s.
\]

 The usual form of Newton's identities \citet{MS001}:
 Let $\omega_1,\ldots,\omega_r$ be indeterminates, and
\[
\sigma(t)=\prod\limits_{j=1}^{r}(1-\omega_jt)=\sum\limits_{j=0}^{r}\sigma_jt^j,
\]
where $\sigma_j$ is an elementary symmetric function of the $\omega_j$, $\sigma_0=1$, and
$\sigma_j=0$ for $j>r$. Define the power sums
\[
P_j=\sum\limits_{i=1}^{r}\omega_i^{j} \ \ \mbox{for all} \ \ j.
\]

\begin{enumerate}
\renewcommand{\labelenumi}{$($\mbox{\roman{enumi}}$)$}
\item
If $P(t)=\sum\limits_{j=1}^{\infty}P_jt^{j}$, then
\[
\sigma(t)P(t)+t\sigma'(t)=0.
\]
\item
By equating coefficients
\[
\begin{array}{ll}
&P_1+\sigma_1=0\\
&P_2+\sigma_1P_1+2\sigma_2=0\\
&\mbox{......................................}\\
&P_r+\sigma_1P_{r-1}+\cdots+\sigma_{r-1}P_1+r\sigma_r=0
\end{array}
\]
\end{enumerate}
and, for $j>r$,
\[
 P_j+\sigma_1P_{j-1}+\cdots+\sigma_{r}P_{j-r}=0.
\]

\subsection{Conjecture 2}\label{III02}

\begin{prop}\label{conj002}
$K_m'=K_m$ for $k=3$ and $3\nmid m$. 
\end{prop}

\begin{pf}
In the following, there are three parts, the first part considers the exponential sum $K_m'$, the second part considers the exponential sum $K_m$. And a comparision between them is given in the third part.

Part I: To investigate the exponential sum $K_m'$ defined in Definition \ref{def01}, we need to consider the trace $\mbox{Tr}_m(f(x))$, which can take only two values $0$ and $1$ that correspond to $1$ and $-1$ respectively in the exponential sum. Let's consider the case that this trace takes value $0$, then according to \citet[Theorem 2.25]{LH001}, $f(x)=y^2+y$ for some $y\in \mbox{GF}(2^m)$. To study the number of times (denoted $n_1$) the trace value is $0$, consider the equation $f(x)=y^2+y$. Now,
\[
 f(x)={{(x^{2^k}+1)x^{2^k}}\over {(x^{2^k}+x)^{2^k+1}}}=y^2+y
\]
after eliminating the denominator, it becomes
\[
(x^{2^k}+1)x^{2^k}=(y^2+y)(x^{2^{2k}}+x^{2^k})(x^{2^k}+x).
\]
Simplifying, we have
\begin{equation}\label{001}
x^{2^k}+1+(y^2+y)(x^{2^{2k}}+x^{2^{2k}-2^k+1}+x^{2^k}+x)=0
\end{equation}
denote the number of its solutions by $n_1'$, then for every solution $(x,y)$, $(x,y+1)$ is also a solution. For $x=1$, equation (\ref{001}) has $2^m$ solutions, but $x=1$ counts only one time in $n_1$, thus
\begin{equation}\label{n1001}
n_1={{n_1'-2^m}\over 2}+1.
\end{equation}

The degree of polynomial equation (\ref{001}) is $2^{2k}+2$, and denote this polynomial by $f_k(x)$.
To study its solutions, let's consider the projective closure of the affine hypersurface defined by it. Define $\bar{f}_k(x,y,z)=z^{\mbox{deg}f_k}f_k({x\over z},{y\over z})$, then $\bar{f}_k$ is a homogeneous polynomial of degree euqal to $\mbox{deg}f_k$. Moreover, $\bar{f}_k(x,y,1)=f_k(x,y)$. We consider
\begin{equation}\label{k001}
\bar{f}_k(x,y,z)=x^{2^k}z^{2^{2k}-2^k+2}+z^{2^{2k}+2}+(y^2+yz)(x^{2^{2k}}+x^{2^{2k}-2^k+1}z^{2^{k}-1}+x^{2^k}z^{2^{2k}-2^k}+xz^{2^{2k}-1})=0.
\end{equation}
The general solutions of above equation may not be so easy to determine, and the case $k=3$ is studied. That is
\begin{equation}\label{002}
\bar{f}_3(x,y,z)=x^{8}z^{58}+z^{66}+(y^2+yz)(x^{64}+x^{57}z^{7}+x^{8}z^{56}+xz^{63})=0
\end{equation}
with the number of projective points denoted by $\bar{n}_1$.
The two points $(1:0:0)$ and $(0:1:0)$ correspond to $z=0$, so they ought to be omitted when counting the number of points for $n_1'$, that is
\begin{equation}\label{n1002}
n_1'=\bar{n}_1-2.
\end{equation}
 Without causing any ambiguity, the indices $m,k$ are omitted .

Equation (\ref{002}) defines a projective variety $\mathcal{P}_1$ of dimension $1$ in a projective $2$-space. The curve has at least one irreducible components over $\mbox{GF}(2)$, corresponding to solutions with two variables equal. This trivial component is deinfed by equation
\[
 x+z=0.
\]
Using Magma \citet{BC001}, the second nontrivial component is singular, absolutely irreducible, and is defined by equation
\[
\begin{array}{rr}
x^{56}y^2 + x^{56}yz + x^{49}y^2z^7 + x^{49}yz^8 + x^{48}y^2z^8 + x^{48}yz^9 +x^{41}y^2z^{15} + x^{41}yz^{16}&\\
+ x^{40}y^2z^{16} + x^{40}yz^{17} + x^{33}y^2z^{23} +x^{33}yz^{24} + x^{32}y^2z^{24} + x^{32}yz^{25}  &\\+x^{25}y^2z^{31} +
x^{25}yz^{32} +x^{24}y^2z^{32} + x^{24}yz^{33} + x^{17}y^2z^{39} + x^{17}yz^{40}+ x^{16}y^2z^{40}&\\  +x^{16}yz^{41} + x^9y^2z^{47} + x^9yz^{48} + x^8y^2z^{48} &\\+
 x^8yz^{49} +xy^2z^{55} + xyz^{56} + z^{58}&=0.
\end{array}
\]
Note that $(0:1:0)$ is the point satisfying both the trivial component and nontrivial component $\tilde{\mathcal{P}}_1$, and there are $2^m$ points of the form $(1:y:1)$ which  belong to the trivial component but not the nontrivial component. Let $\tilde{n}_1$ be the number of point on $\tilde{\mathcal{P}}_1$, then
\begin{equation}\label{n1003}
\bar{n}_1=\tilde{n}_1+2^m.
\end{equation}

Magma shows that the projective variety $\tilde{\mathcal{P}}_1$ has genus 31 \citet{BC001}. As in \citet{BM001}, using Magma one can compute the zeta function of a variety. It is important to note that Magma actually computes the zeta function of the nonsingular model.
Since the curve $\tilde{\mathcal{P}}_1$ is singular, we need to consider it carefully as what has been done in \citet{BM001}, where the singular points are analyzed according to whether $m$ is even or odd, see Remark \ref{rsnm001}.

Let $\hat{\mathcal{P}}_1$ be the nonsingular model of $\tilde{\mathcal{P}}_1$, Magma computes the zeta function  of $\hat{\mathcal{P}}_1$ to be
\[
\begin{array}{lll}
\mathcal{Z}_1(t)=&(4096t^{24} + 512t^{21} + 128t^{18} - 128t^{15} - 16t^{12} - 16t^9 + 2t^6 + t^3+1)\\
&(32768t^{30} - 4096t^{27} - 2048t^{24} - 1280t^{21} + 256t^{18} + 96t^{15} +32t^{12} - 20t^9 - 4t^6 - t^3 + 1)\\
&(2t^2 + t + 1)(8t^6 + 2t^3 + 1)/(2t^2 - 3t + 1).
\end{array}
\]
Denote the number of points on $\hat{\mathcal{P}}_1$ by $\hat{n}_1$, then
\begin{equation}\label{n1004}
\tilde{n}_1=\hat{n}_1-S_m^1
\end{equation}
 where $S_m^1=2^{1+\delta_{3,m}}$, $\delta_{3,m}=0$ when $3\nmid m$ and $2$ otherwise.

Combing equations (\ref{n1001},\ref{n1002},\ref{n1003},\ref{n1004}),
\begin{equation}\label{n1005}
n_1={{\hat{n}_1-S_m^1}\over 2}.
\end{equation}

For
\[
\mathcal{Z}_1(t)={{L_1(t)}\over{(1-t)(1-2t)}}
\]
assume that ${L_1(t)}=\prod\limits_{j=1}^{62}(1-\kappa_jt)$, then the number of points over $\mbox{GF}(2^m)$ on $\hat{\mathcal{P}}_1$ is
\[
\hat{n}_1=2^m+1-\sum\limits_{j=1}^{62}\kappa_j^{m}.
\]

Part II: As to the exponential sum $K_m$, we cnosider the number ($n_2'$) of solutions of the following polynomial equation
\[
x+x^{-1}=y^2+y.
\]
Eliminating the denominator and simplifying,  
\[
g(x,y)=x^2+1+(y^2+y)x=0
\]
and the projective closure $\mathcal{P}_2$ of the affine space is defined by
\[
\bar{g}(x,y,z)=x^2z+z^3+(y^2+yz)x
\]
which is irreducible, nonsingular, and with genus $1$.

Using Magma, one can compute the zeta function,
\[
 \mathcal{Z}_2(t)={{L_2(t)}\over{(1-t)(1-2t)}}={{2t^2 + t + 1}\over {2t^2 - 3t + 1}}
\]
where $L_2(t)=2t^2 + t + 1=(1-\pi_1t)(1-\pi_2t)$, then the number of points over $\mbox{GF}(2^m)$ on $\mathcal{P}_2$ is
\[
\bar{n}_2=2^m+1-\pi_1^m-\pi_2^m.
\]

Let $n_2$ denote the number of times that $\mbox{Tr}_m(x+x^{-1})=0$ when $x\in \mbox{GF}(2^m)^*$. Since $(1:0:0)$ and $(0:1:0)$ ought to be omitted when counting the number of points on $\mathcal{P}_2$,  
\begin{equation}\label{n2001}
n_2={n_2'\over 2}={{\bar{n}_2-2}\over 2}.
\end{equation}

Part III: To compare the number of points on $\hat{\mathcal{P}}_1$ and the number of points on $\mathcal{P}_2$, let's consider the polynomials $L_1(t)$ and $L_2(t)$. Divide $L_1(t)$ by $L_2(t)$  
\[
\begin{array}{lll}
L_1'(t)=&(4096t^{24} + 512t^{21} + 128t^{18} - 128t^{15} - 16t^{12} - 16t^9 + 2t^6 + t^3+1)\\
&(32768t^{30} - 4096t^{27} - 2048t^{24} - 1280t^{21} + 256t^{18} + 96t^{15} +32t^{12} - 20t^9 - 4t^6 - t^3 + 1)\\
&(8t^6 + 2t^3 + 1)
\end{array}
\]
where we assume ${L_1'(t)}=\prod\limits_{j=1}^{60}(1-\kappa_jt)$. Then
\[
d_m=\bar{n}_2-\hat{n}_1=\sum\limits_{j=1}^{60}\kappa_j^{m}.
\]

Now $L_1'(t)$ can be expanded as follows
\begin{equation}\label{dl002}
\begin{array}{lll}
L_1'(t)=&1073741824t^{60} + 268435456t^{57} + 83886080t^{54} - 100663296t^{51} -27262976t^{48} -9437184t^{45}  \\
& + 4784128t^{42} + 1179648t^{39} + 573440t^{36} -212992t^{33} - 41984t^{30} - 26624t^{27} + 8960t^{24} \\
&+ 2304t^{21} + 1168t^{18} -288t^{15} - 104t^{12} - 48t^9 + 5t^6 + 2t^3 + 1.
\end{array}
\end{equation}
Denote the coefficient of $L_1'(t)$ by $\sigma_j$ which corresponds to $t^j$, $j=0,1,2,\ldots, 60$.
Direct verification, $d_m$ can be found to be zero for those $m\leq 60$ and not divisible by $3$.

Let's consider the other cases. Suppose this holds for $m\leq i$, then for $m=i+1=3j+1$,  
\[
\begin{array}{lll}
d_{m}&=-(\sum\limits_{s=1}^{60}\sigma_{s}d_{3j+1-s})\\
    &=-(\sigma_1d_{3j}+\sigma_2d_{3j-1}+\sigma_3d_{3j-2}+\sigma_4d_{3j-3}+\cdots+\sigma_{58}d_{3j-57}+\sigma_{59}d_{3j-58}+\sigma_{60}d_{3j-59})
\end{array}
\]
which is zero by noting the coefficients of (\ref{dl002}) and the assumption for the $m\leq i$ case.
Similarly, for $m=i+2=3j+2$
\[
\begin{array}{lll}
d_{m}&=-(\sum\limits_{s=1}^{60}\sigma_{s}d_{3j+2-s})\\
    &=-(\sigma_1d_{3j+1}+\sigma_2d_{3j}+\sigma_3d_{3j-1}+\sigma_4d_{3j-2}+\cdots+\sigma_{58}d_{3j-56}+\sigma_{59}d_{3j-57}+\sigma_{60}d_{3j-58})
\end{array}
\]
which is also zero. By induction, $d_m$ is zero for all $m$ that cann't be divided by $3$. That is
\begin{equation}\label{n12001}
\hat{n}_1=\bar{n}_2 \ \ \mbox{for} \ \ 3\nmid m.
\end{equation}

From equations (\ref{n1005},\ref{n2001},\ref{n12001}),  
\begin{equation}\label{n12002}
n_1=n_2
\end{equation}
when $3\nmid m$. That is $K_m'=K_m$.
\end{pf}

\begin{rem}\label{rsnm001}
In fact, the zeta function of the singular curve $\tilde{\mathcal{P}}_1$ can be found from the zeta function of the nonsingular model $\hat{\mathcal{P}}_1$ 
\[
\begin{array}{lll}
 &(4096t^{24} + 512t^{21} + 128t^{18} - 128t^{15} - 16t^{12} - 16t^9 + 2t^6 + t^3+1)\\
&(32768t^{30} - 4096t^{27} - 2048t^{24} - 1280t^{21} + 256t^{18} + 96t^{15} +32t^{12} - 20t^9 - 4t^6 - t^3 + 1)\\
&(2t^2 + t + 1)(8t^6 + 2t^3 + 1)(t^2+t+1)^2(t-1)^4/(2t^2 - 3t + 1).
\end{array}
\]
As in \citet{BM001}, we work with the nonsingular, irreducible model. 
\end{rem}

\subsection{Conjecture 1}

\begin{prop}\label{conj001}
$G_m^{(3)}=G_m$ for $3\nmid m$. 
\end{prop}

\begin{pf}
In the following, exponential sums $G_m^{(3)}$ and $G_m^{(1)}$ are analyzed separately, and then compared.

For the exponential sum $G_m^{(k)}$, let $n_3$ denote the number of times $\mbox{Tr}_m(x^{2^k+1}+x^{-1})=0$ ($x\in \mbox{GF}(2^m)^*$). We consider the following equation
\[
x^{2^k+1}+x^{-1}=y^2+y,
\]
that is
\[
f_k(x,y)=x^{2^k+2}+(y^2+y)x+1=0
\]
with the number of solutions denoted by $n_3'$, and $n_3=n_3'/2$.
The projective closure is
\[
\bar{f}_k(x,y,z)=x^{2^k+2}+(y^2+yz)xz^{2^k-1}+z^{2^k+2}.
\]
For $k=3$, 
\[
\bar{f}_3(x,y,z)=x^{10}+(y^2+yz)xz^{7}+z^{10}
\]
which is irreducible and singular, and the number of solutions $\bar{n}_3$. Note that $(0:1:0)$ is the point that should be omitted when counting $\bar{n}_3$, thus $n_3'=\bar{n}_3-1$.
Denote the projective variety by $\mathcal{P}_3$ and the nonsingular model by $\hat{\mathcal{P}}_3$ (with number of points $\hat{n}_3$). Using Magma the zeta function corresponding to $\hat{\mathcal{P}}_3$ can be found to be
\[
\mathcal{Z}_3={{(2t^2 + 2t + 1)(4t^4 - 4t^3 + 2t^2 - 2t + 1)(4t^4 + 2t^3 + t + 1)}\over {2t^2 - 3t + 1}}.
\]
Let
\[
{L_3(t)}=(2t^2 + 2t + 1)(4t^4 - 4t^3 + 2t^2 - 2t + 1)(4t^4 + 2t^3 + t + 1)=\prod\limits_{j=1}^{10}(1-\kappa_jt).
\]
Then the number of points over $\mbox{GF}(2^m)$ on $\hat{\mathcal{P}}_3$ is
\[
\hat{n}_3=2^m+1-\sum\limits_{j=1}^{10}\kappa_j^m.
\]
 It can be verified that the only singular point of $\mathcal{P}_3$ is $(x:y:z)=(0:1:0)$. Thus $\bar{n}_3=\hat{n}_3-1$, and
\[
n_3={{\hat{n}_3-2}\over 2}.
\]

As for the exponential sum $G_m^{(1)}$, let $n_4$ denote the number of times $\mbox{Tr}_m(x^3+x^{-1})=0$ when $x\in \mbox{GF}(2^m)^*$. We need to consider
\[
x^3+x^{-1}=y^2+y
\]
that is $g(x,y)=x^4+(y^2+y)x+1=0$ with number of solutions $n_4'$, and $n_4={n_4'\over 2}$. The projective closure $\mathcal{P}_4$ is defined by
\[
\bar{g}(x,y,z)=x^4 + xy^2z + xyz^2 + z^4
\]
which is irreducible, singular and number of points $\bar{n}_4$. With an exceptional point $(0:1:0)$, $n_4'=\bar{n}_4-1$. Also, the zeta function of the nonsingular model ($\hat{\mathcal{P}}_4$, with number of points $\hat{n}_4$) can be calculated  
\[
\mathcal{Z}_4={{4 t^4 + 2 t^3 + t + 1}\over {2 t^2 - 3 t + 1}}.
\]

The singular point of $\mathcal{P}_4$ is $(x:y:z)=(0:1:0)$, then $\bar{n}_4=\hat{n}_4-1$, and
\[
n_4={{\hat{n}_4-2}\over 2}.
\]

Suppose ${L_4(t)}={4 t^4 + 2 t^3 + t + 1}=\prod\limits_{j=7}^{10}(1-\kappa_jt)$, then the number of points over $\mbox{GF}(2^m)$ on $\hat{\mathcal{P}}_4$ is
\[
\hat{n}_4=2^m+1-\sum\limits_{j=7}^{10}\kappa_j^m.
\]

Note that ${L_4(t)}$ is a factor of ${L_3(t)}$,  
\[
{L_3'(t)}={{L_3(t)}\over {L_4(t)}}=(2t^2 + 2t + 1)(4t^4 - 4t^3 + 2t^2 - 2t + 1)=8t^6 - 4t^3 + 1
\]
thus,
\[
d_m=\hat{n}_4-\hat{n}_3=\sum\limits_{j=1}^{6}\kappa_j^m.
\]

Since the coefficient $\sigma_j$ (corresponding to $t^j, j=0,1,2,\ldots,6$) of $L_3'(t)$ is zero when $3\nmid j$, it can be verified that $d_m=0$ for $m$ that is not divisible by $3$. Thus $G_m^{3}=G_m$. 
\end{pf}

\begin{rem}
Note that in Propositions \ref{conj002} and \ref{conj001},  the assumption $m$ is odd is not needed.
\end{rem}

Combing Corollary \ref{cor001} and Propositions \ref{conj002}, \ref{conj001} we have
\begin{thm}\label{CD001}
The cross-correlation distribution is completely determined for the decimation $d=65/9$, when $3\nmid m$ is odd, by Theorem \ref{thm001} and $A_1=2^m+1+3G_m-2K_m-2C_m$. The distribution is the same as that of $d=5/3, 17/5$.
\end{thm}

\begin{exmp}
Let $\alpha$ be a primtive element of the finite field $\mbox{GF}(2^m)$. For the weight distributions of the cyclic codes with two nonzeros $\alpha^{-(2^{2\cdot 1}+1)}$ and $\alpha^{-({2^{1}+1})}$, there are the following results in \citet{BM001}. For $m=7$, the corresponding cyclic code $\mathcal{C}_1$ has three nonzero weights and
\[
A_{64}=8255, A_{56}=4572, A_{72}=3556.
\]
For $m=11$, the corresponding cyclic code $\mathcal{C}_2$ has five nonzero weights and
\[
A_{1024}=2368379, A_{992}=900680, A_{1056}=835176, A_{960}=45034, A_{1088}=45034.
\]

Using Matlab, it can be found that the cyclic code with nonzeros $\alpha^{-(2^{2\cdot 3}+1)}$ and $\alpha^{-({2^{3}+1})}$ has the same weight distribution as $\mathcal{C}_1$ for $m=7$, and $\mathcal{C}_2$ for $m=11$. This confirms the result of Theorem \ref{CD001}.

\end{exmp}

\section{Conclusion} \label{secVII}

 Two conjetures are presented in \citet{JHK001}, and solved for the cases $k=1,2$. In this correspondence, we study the case $k=3$, as a result of which the five-valued cross-correlation distribution is determined for another case.

\begin{ack}
The authors would like to thank  the anonymous reviewers for many helpful
comments. In particular,
we would like to thank N. Sutherland for the help with the use of Magma.
\end{ack}

\end{document}